\documentclass[10pt]{article}
\usepackage{graphics}
\usepackage{latexsym}
\usepackage{amssymb}
\usepackage{amsmath}
\usepackage{amscd}
\usepackage{graphicx}
\usepackage{textcomp}
\usepackage{colortbl}
\usepackage{subcaption}
\usepackage{hyperref}
\usepackage{multirow}
\usepackage{lipsum}
\usepackage[left=1.5cm,top=1.5cm,right=1.5cm,bottom=1.5cm]{geometry}
\begin{document}
\begin{center}
\large{\bf{Recovering $\Lambda$CDM Model From a Cosmographic Study}} \\
\vspace{10mm}
\normalsize{Hassan Amirhashchi$^{1}$, Soroush Amirhashchi$^{2}$\\
\vspace{5mm}
\normalsize{$^{1,2}$Department of Physics, Mahshahr Branch, Islamic Azad University,  Mahshahr, Iran \\
E-mail:$^{1}$h.amirhashchi@mhriau.ac.ir},~~~$^{2}$ soroush.amirhashchi@gmail.com} \\
\end{center}
\begin{abstract}
	
Using the mathematical definitions of deceleration and jerk parameters we obtain a general differential equation for squared Hubble parameter. For a constant jerk, this differential equation leads to an exact function for Hubble parameter. By the aid of this exact Hubble function we can exactly reconstruct any other cosmographic parameters. We also obtained a general function for transition redshift as well as spacetime curvature. Our derived functions clearly impose a lower limit on the jerk parameter which is $j_{min}\geq-0.125$. Moreover, we found that the jerk parameter indicates the geometry of the spacetime i.e any deviation from $j=1$ imply to a non-flat spacetime. In other word $j\neq 1$ reefers to a dynamical, time varying, dark energy. From obtained Hubble function we recover the analogue of $\Lambda$CDM model. To constrain cosmographic parameters as well as transition redshift and spacetime curvature of the recovered $\Lambda$CDM model, we used Metropolis-Hasting algorithm to perform Monte Carlo Markov Chain analysis by using observational Hubble data obtained from cosmic chronometric (CC) technique,  BAO data, Pantheon compilation of Supernovae type Ia, and their joint combination. The only free parameters are $H$, $A(\Omega_{m})$ and $j$. From joint analysis we obtained $H_{0}=69.9\pm 1.7$, $A(\sim\Omega_{0m})=0.279^{+0.013}_{-0.017}$, $B(\sim\Omega_{0X})=0.721^{+0.017}_{-0.013}$, $j_{0}=1.038^{+0.061}_{-0.023}$ and $z_{t}=0.706^{+0.031}_{-0.034}$.
\end{abstract}
\smallskip
{\it Keywords}: Cosmography; Hubble Rate; Deceleration Parameter; Jerk Parameter\\
PACS Nos: 98.80.Es, 98.80.-k, 95.36.+x, 98.80.Jk
\section{Introduction}
\label{sec:1}
Since 1998 it has been revealed that the cosmic expansion is speeding up \cite{ref1,ref2,ref3,ref4}. In the context of general theory of relativity (GR) the existence of an exotic fluid called \textgravedbl dark energy (DE)\textacutedbl with negative pressure is considered as the source caused present universe accelerated expansion \cite{ref5,ref6,ref7}. It is also possible to study this cosmic acceleration in the context of modified gravity which is a generalization of general relativity \cite{ref8,ref9}. Moreover one may deal with this problem by considering violation of cosmological principle i.e assuming that the spacetime metric is inhomogeneous. Despite of about two decades effort, we still could not propose a realistic cosmological model in order to describe the present day cosmic accelerated expansion precisely. For example, although $\Lambda$CDM model \cite{ref10} excellently fits all observational data (for example see \cite{ref11}) but considering cosmological constant $\Lambda$ as dark energy encounters to coincidence and fine-tuning problems \cite{ref12,ref13}. We know that the modern cosmology is based on the Friedmann equations, nevertheless, it is interesting to study universe through a kinematic approach rather than dynamical one (Einstein equations). This model-independent approach is called \textgravedbl Cosmography\textacutedbl or cosmo-kinetics . It is worth noting that in this purely kinematic approach all the derived quantities are also model-independent. Cosmography was firstly introduced by Weinberg \cite{ref14} in 2008 and  later extended by Visser \cite{ref15}. The cornerstone of cosmography is to expand some observables such as the Hubble parameter(or equivalently the scale factor) into power series, and directly relating cosmological parameters to these observable quantities.\\
However, in practice, the cosmography study confronts to two serious problems with observational data. First of all, as shown in Ref \cite{ref17}, the Taylor expansion fails to reach convergence  at redshift $z>1$. This, of course an important shortcoming as many observational probes such as supernovae type Ia (SNIa) and cosmic microwave background (CMB) compilations can span the redshift region up to $z\sim 1.3$ and $z\sim 1100$ respectively. This problem could be overcome by definition of an improved redshift parametrization such as $y=1/(1+z)$ \cite{ref17,ref18,ref19}. The second one is the fact that in cosmography we use a finite Taylor series truncation which represent an approximation of the exact function and hence leads to worse estimations. Note that, taking more terms of Taylor series gives raise to more precise approximation but higher errors. Therefore as mentioned in Ref \cite{ref19} \textgravedbl cosmography is in the dilemma between accuracy and precision\textacutedbl. For recent cosmography studies in the context for GR and Modified GR Reader is advised to see Refs \cite{ref19,ref20,ref21,ref22,ref23,ref24,ref25,ref26,ref27,ref28} an \cite{ref29,ref30} respectively. Very recently it has been shown that using Weighted Function Regression method \cite{ref31} improves the usual cosmographic approach by automatically implementing Occam’s razor criterion \cite{ref32}.\\

As the source of all above mentioned problems lies in the Taylor expansion of the Hubble (scale factor) or luminosity distance, in this paper, in contrast with usual cosmography, we do not expand any of these parameters and instead we try to find an exact function for Hubble parameter on the bases of kinematic parameters of the universe. To do so, we mix the definition of deceleration and jerk parameters which in turn gives a second order differential equation for squared Hubble parameter. Considering $j$ as a slowly varying (constant) quantity, this differential equation leads to an exact Hubble parameter as a function of jerk parameter (see sec~\ref{sec:3}). Through this Hubble parameter we reconstruct all other cosmographic parameters (CS) exactly. Then we use observational Hubble data (OHD) in the redshift range $0.07\leq z\leq 1.965$ \cite{ref33}, Pantheon compilation containing 1048 SNIa apparent magnitude measurements over the redshift range of $0.01 < z < 2.3$ \cite{ref34}, BAO dataset (containing ten data \cite{ref35}), and their joint combination data to constrain CS parameters. We compare some of our results by those obtained in Refs \cite{ref18,ref20} and \cite{ref22}. This paper is organized as follows. In Sec~ \ref{sec:2} we briefly discuss cosmography and introduce CS parameters up to the fifth derivative of the scale factor. In Sec~ \ref{sec:3} we derive a general deferential equation for squared Hubble parameter an reconstruct all CS parameter from it. Subsec~ \ref{subsec:3.1} deals with the derivation (recovering) $\Lambda$CDM model from our almost-general Hubble solution and show that how the spacetime geometry is connected to the jerk parameter. We summary the computational technique we have used to fit CS parameters to data by a numerical MCMC analysis in Sec~ \ref{sec:4}. Sec~ \ref{sec:5} deals with the results of our fits to data. In Subsec~ \ref{subsec:5.1} we derive a general redshift and constrain it over data. Finally, we summarize our findings and conclusions in Sec~ \ref{sec:6}.

\section{Cosmography}
\label{sec:2}
In this section we shall briefly describe cosmography which may start from Taylor series of scale factor.\\
Taylor expansion of the scale factor $a(t)$ around the current time $t_{0}$ gives

\begin{equation}
\label{eq1}
a(t)=a_{0}\left[1+ \sum_{n=1}^{\infty}\frac{1}{n!}\dfrac{d^{n}a}{dt^{n}}(t-t_{0})^{n}\right].
\end{equation}

Without loss of generality we can assume $a_{0}=1$, where the constant $a_{0}$ is the current value of the scale factor. The most important cosmographic series terms i.e the Hubble, deceleration, jerk, snap, and lerk parameters are \cite{ref16,ref32,ref36}

\begin{subequations}
\label{eq2}	
\begin{align}
H&=\dfrac{1}{a}\dfrac{da}{dt},\label{eq2a}\\
q&=-\dfrac{1}{aH^{2}}\dfrac{d^{2}a}{dt^{2}},\label{eq2b}\\
j&=\dfrac{1}{aH^{3}}\dfrac{d^{3}a}{dt^{3}},\label{eq2c}\\
s&=\dfrac{1}{aH^{4}}\dfrac{d^{4}a}{dt^{4}}\label{eq2d}\\
l&=\dfrac{1}{aH^{5}}\dfrac{d^{5}a}{dt^{5}}\label{eq2e},
\end{align}
\end{subequations}

respectively. As noted in Ref \cite{ref29} the first three CS parameters i.e the Hubble rate $H$, deceleration parameter $q$ and its first derivative with respect to the cosmic time ( or redshift) $j$  are sufficient to determine the overall kinematics of the Universe. However, at the current time the deceleration parameter is restricted as $-1\leq q_{0}<0$ which in turn impose $j_{0}>0$. For $\Lambda$CDM model $j=1$ at any time \cite{ref37}.

Since $a=1/(1+z)$, from eqs (\ref{eq2}), we can find the derivatives of the Hubble parameter with respect to $z$ as \cite{ref19}

\begin{subequations}
\label{eq3}	
\begin{align}
\frac{d{H}}{d{z}}&=\dfrac{1+q}{1+z}H,\label{eq3a},\\
\frac{d^{2}{H}}{d^{2}{z}}&=\dfrac{j-q^{2}}{(1+z)^{2}}H,\\
\frac{d^{3}{H}}{d^{3}{z}}&=\dfrac{H}{(1+z)^{3}}(3q^{2}+3q^{3}-4qj-s)\\
\begin{split}
\frac{d^{4}{H}}{d^{4}{z}}&=\dfrac{H}{(1+z)^{4}}(-12q^{2}-24q^{3}-15q^{4}+32qj+25q^{2}j\\&+7qs+12j-4j^{2}+8s+l).
\end{split}
\end{align}
\end{subequations}

Using eqs (\ref{eq3}) in the Taylor expansion of the Hubble parameter around $z=0$ one can obtain

\begin{align}
\label{eq4}
\begin{split}
H(z)&=H_{0}+\left.\frac{dH}{dz}\right\vert_{z=0}z+\left.\frac{1}{2}\dfrac{d^{2}H}{dz^{2}}\right\vert_{z=0}z^{2}+\left.\frac{1}{6}\dfrac{d^{3}H}{dz^{3}}\right\vert_{z=0}z^{3}+ O(z^{4})\\
&\qquad =H_{0}\biggl[1+(1+q_{0})z+\frac{1}{2}(j_{0}-q_{0}^{2})z^{2}\\
&\qquad +\dfrac{1}{6}(3q_{0}^{2}+3q_{0}^{3}-4q_{0}j_{0}-3j_{0}-s_{0})z^{3}+O(z^{4})\biggr],
\end{split}
\end{align}
In other hand, the cosmographic version of the luminosity distance can be conveniently expressed as

\begin{align}
\label{eq5}
d_{L}(z) &= \dfrac{cz}{H_{0}}\biggl[1+\dfrac{1}{2}(1-q_{0}z)-\frac{1}{6}(1-q_{0}-3q_{0}^{2}+j_{0})z^{2}\nonumber \\
&+\dfrac{1}{24}(2-2q_{0}-15q_{0}^{2}-15q_{0}^{3}+5j_{0}+10j_{0}q_{0}+s_{0})z^{3}+O(z^{4})\biggr]
\end{align}
where the subscript \textquotedblleft0\textquotedblright indicates the present values of the cosmographic parameters.\\

Here we again emphasize that there are two main problems arising in the context of
cosmography when using eqs (\ref{eq4}) and (\ref{eq5}) to constrain CS parameters. In fact, it has already been shown that expanding around $z \sim 0$ gives raise to the divergence of Taylor series at $z\geq0$. This problem could be limited by transferring $z$ to $z/(1+z$). The second problem is the truncation of the series one may use in analysis, an approximation of the exact function, which may leads to the possible misleading results. Although, this problem can be alleviated by going to higher terms in the expansion, but adding any new term means introducing a new parameter that must be estimated. This indeed increases the divergences of the analysis. Moreover, since this method is based on the Taylor expansion of the scale factor, cosmography is restricted in the scope of Friedmann-Robertson-Walker metric. Hence it is interesting to find a more common kinematic approach which is applicable in other spacetimes posses inhomogeneous properties.\\ In next section we will derive an almost-general model independent solution from which one can reconstruct any CS parameter without any limitation and problems arising in usual cosmography.

\section{An Exact Cosmographic Solution}
\label{sec:3}
In this section we find an exact analytical expressions between the Hubble and jerk parameters. It is interesting to note that in \cite{ref38} an almost the same expression has been found by considering special parametrization for the jerk parameter, but this solution seems to be wrong as the authors consider a minus sign in the definition of jerk parameter (see eq (5) of this reference). This mistake affects other results. In what follows, first we derive a general differential equation which could be used for any parametrization of the jerk parameter and then we solve it for $j(z)=j_{0}$. It is worth to mention that, from theoretical point of view, while $j$ is constant for $\Lambda$CDM model for most other models of dark energy evolves with time. However, very recently G\'{o}mez-Valent \cite{ref32} obtained a reconstructed jerk as a function of the redshift which is completely compatible with a constant value and, in particular, with the $\Lambda$CDM value $j=1$. Consequently, because the data is not constraining enough at the moment to characterize the time evolution of the jerk, it will be licit to consider jerk as a constant parameter (see ref \cite{ref32} for more details).\\

It is possible to obtain the derivatives of the squared Hubble parameter with respect to $z$ as follows \cite{ref39}

\begin{subequations}
\label{eq6}	
\begin{align}
\frac{d{H^{2}}}{d{z}}&=\dfrac{2H^{2}}{1+z}(1+q),\label{eq6a}\\
\frac{d^{2}{H^{2}}}{d{z^{2}}}&=\dfrac{2H^{2}}{(1+z)^{2}}(1+2q+j),\label{eq6b}\\
\frac{d^{3}{H^{2}}}{d{z^{3}}}&=\dfrac{2H^{2}}{(1+z)^{3}}(-qj-s)\label{eq6c}\\
\frac{d^{4}{H^{2}}}{d{z^{4}}}&=\dfrac{2H^{2}}{(1+z)^{4}}(4qj+3qs+3q^{2}j-j^{2}+4s+l)\label{eq6d}.
\end{align}
\end{subequations}
Combining eqs (\ref{eq6a}) and (\ref{eq6b}) we find the following differential equation for squared Hubble parameter

\begin{equation}
\label{eq7}
\dfrac{1}{2}(1+z)^{2}\dfrac{(H^{2})''}{H^{2}}-(1+z)\dfrac{(H^{2})'}{H^{2}}+(1-j)=0,
\end{equation}

which in turn, for constant $j$ \footnote{Note that one can parameterize jerk parameter as $j=A+B/(1+z)$ to find a solution for non-constant $j$ which is in consistent with the $\Lambda$CDM scenario at current time.}, gives the following general solution for the Hubble parameter

\begin{equation}
\label{eq8}
h(z)^{2}=\left[A(1+z)^{\frac{3+\sqrt{1+8j}}{2}}+B(1+z)^{\frac{3-\sqrt{1+8j}}{2}}\right],
\end{equation}

where $h(z)=H(z)/H_{0}$ . Requiring the consistency of (\ref{eq8}) at $z=0$ gives

\begin{equation}
\label{eq9}
A+B=1.
\end{equation}
Obviously eq (\ref{eq8}) is a two parameter model of $j$ and $A$. Generally, this solution corresponds to a cosmological model without radiation component \footnote{One can put eqs (\ref{eq6a}) and (\ref{eq6b}) in eq (\ref{eq6c}) to find a similar differential equation as eq (\ref{eq7}) in term of snap parameter. Doing so, probably, recovers radiation term.}. Note that, as we have shown in next section there is  degeneracy between curvature $\Omega_{k}$ and $j$. It is worth to mention that for spatially flat cosmological constant dark energy model the jerk parameter is $j (z) = 1$. Therefore, jerk parameter has been a traditional tool to test the spatially flat $\Lambda$CDM model. From eq (\ref{eq8}) it is clear that considering $j=1$ recovers flat $\Lambda$CDM model. Hence, in this case, one can consider $A$ and $B$ as matter and dark energy density parameters respectively.\\

Using eq (\ref{eq8}) in eqs (\ref{eq6}) we can reconstruct cosmographic parameters as

\begin{subequations}
\label{eq10}	
\begin{align}
q=&-1+\dfrac{1}{2h^{2}(z)}\biggl[AX(1+z)^{X}+BY(1+z)^{Y}\biggr],\label{eq10a}\\
j=&1-\dfrac{1}{2h^{2}(z)}\biggl[A\prod_{i=0}^{1}(X-i)(1+z)^{X} \nonumber \\
&+B\prod_{i=0}^{1}(Y-i)(1+z)^{Y}\biggr],\label{eq10b} \\
s=&-qj-\dfrac{1}{2h^{2}(z)}\biggl[A\prod_{i=0}^{2}(X-i)(1+z)^{X} \nonumber \\
&+B\prod_{i=0}^{2}(Y-i)(1+z)^{Y}\biggr] \label{eq10c} \\
l=&-(4qj+3qs+3q^{2}j-j^{2}+4s) \nonumber \\
&-\dfrac{1}{2h^{2}(z)}\biggl[A\prod_{i=0}^{3}(X-i)(1+z)^{X} \nonumber \\
&+B\prod_{i=0}^{3}(Y-i)(1+z)^{Y}\biggr]\label{eq10d},
\end{align}
\end{subequations}
where $X=\frac{3+\sqrt{1+8j}}{2}$, and $Y=\frac{3-\sqrt{1+8j}}{2}$. In the same manner one can reconstruct any other CS parameters.\\
Although one can use the following dimensionless Hubble parameter
\begin{align}
\label{11}
h^{2}(z)&=-\dfrac{1}{2(4qj+3qs+3q^{2}j-j^{2}+4s+l)}\nonumber \\&\left[A\prod_{i=0}^{3}(X-i)(1+z)^{X}+B\prod_{i=0}^{3}(Y-i)(1+z)^{Y}\right],
\end{align}
to put constrain on the cosmographic parameters, but, in view of eqs (\ref{eq8}) \& (\ref{eq9}), the only free parameters are $H_{0}, A$, and $ j$.\\
It is interesting to mention that using eqs (\ref{eq2b}) and (\ref{eq2c}) one can write $j$ in terms of $q$ as follows
\begin{equation}
\label{eq12} j(z)=q(z)(2q(z)+1)+(1+z)\frac{d(z)}{dz}.
\end{equation}
for constant jerk above equations gives the following exact function 
\begin{equation}
\label{eq13} q(z)=\frac{1}{4}\tanh\left[\frac{1}{2}\ln(1+z)\sqrt{1+8j}+\frac{1}{2}c\sqrt{1+8j}\right]\sqrt{1+8j}-\frac{1}{4},
\end{equation}
which in turn for $\Lambda$CDM model leads to
\begin{equation}
\label{eq14} q(z)=\frac{cf(z)+1}{2cf(z)-1};~~~~ f(z)=z^{3}+z^{2}+z+1.
\end{equation}
Since at current time, $z=0; f(z)|_{z=0}=1$, the deceleration parameter is negative and hence from eq (\ref{eq14}) we obtain
\begin{equation}
\label{eq15} q(z)|_{z=0}=\frac{c+1}{2c-1}<0\Rightarrow -1<c<0.5.
\end{equation}
Note that in above equations $c$, is an integrating constant.
\subsection{Recovering $\Lambda$CDM Model}
\label{subsec:3.1}
Recently, Mukherjee \&  Banerjee \cite{ref40} have shown that a $\Lambda$CDM model is favoured where j is allowed to be a function of $z$. Later on, they also show the same result for constant jerk \cite{ref41}. In this section we consider that cosmological constant plays the role of dark energy. Therefore, one can write Friedmann equation as
\begin{equation}
\label{eq16} \dot{a}^{2}+k=\dfrac{1}{3}\rho a^{2} + \dfrac{1}{3}\Lambda a^{2}.
\end{equation}
Taking $\dddot{a}$ we obtain 
\begin{equation}
\label{eq17} j=\dfrac{\dddot{a}}{aH^{3}}=\Omega_{m}a^{-3}+\Omega_{\Lambda}a
\end{equation}
Since, in general, $\Omega_{m}+\Omega_{\Lambda}+\Omega_{k}=1$, we can rewrite above equation as follows
\begin{equation}
\label{eq18} j=1+ \Omega_{m}(a^{-3}-a)-\Omega_{k}a.
\end{equation}
Therefore, we can evaluate the current value of curvature parameter $\Omega_{k}$ only in term of jerk parameter as
\begin{equation}
\label{eq19} \Omega_{0k}=1-j_{0}
\end{equation}
This equation clearly shows that any deviation from $j=1$ is an evidence of non-flat universe. {\bf Therefore we may argue that for all models consider cosmological constant as dark energy, the geometry of spacetime must be flat}. Moreover, when $j\neq 1$ a time varying dark energy is responsible for current Universe accelerating expansion. It is worth nothing that as mentioned in ref \cite{ref41}, if cosmological constant caused the current cosmic acceleration, then there is no any interaction between cosmic dark components.\\

Substituting eq (\ref{eq19}) in eq (\ref{eq8}), we can easily recover the analogue of $\Lambda$CDM model (without radiation) as

\begin{equation}
\label{eq20}
h(z)^{2}=\left[A(1+z)^{\frac{3+\sqrt{\Omega_{k}+9j}}{2}}+B(1+z)^{\frac{3-\sqrt{\Omega_{k}+9j}}{2}}\right],
\end{equation}

which clearly shows degeneracy between $\Omega_{k}$ and $j$. In view of eq (\ref{eq20}), we may also consider A as matter density ($\Omega_{m}$) and B as dark energy density ($\Omega_{X}$) if our estimations indicate  $\Omega_{k} \sim 0$, otherwise we consider $B=\Omega_{X}+\Omega_{k}$.\\

In the next section we will use observational Hubble data (OHD), BAO data and Pantheon compilation and their joint combination to constrain cosmographic $\Lambda$CDM model with following parameters space

\begin{equation}
\label{eq21}
{\bf\Theta}= \{H_{0}, q_{0}, j_{0}, s_{0}, l_{0}, z_{t}, \Omega_{k}, A, B\},
\end{equation}

where $z_{t}$ is the deceleration-acceleration transition redshif (see subsec.~ \ref{subsec:5.1} for more details). Note that $1+8j$ must be greater or equal to zero, this in fact imposes a certain lower limit on jerk parameter as $j\geq -0.125$. We have to consider this point in our estimations.
 
\section{Data and Method}
\label{sec:4}
In this section we briefly describe the astronomical data and the statistical method we have been used to constrain parameter set (\ref{eq21}).

{\bf Type Ia Supernovae:} We adopt the Pantheon compilation \cite{ref34} containing 1048 SNIa apparent magnitude measurements over the redshift range of $0.01 < z < 2.3$ , which includes 276 SNIa ($0.03 < z < 0.65$) discovered by the Pan-STARRS1 Medium Deep Survey and SNIa distance estimates from SDSS, SNLS and low-zHST samples. It is also possible to use the JLA dataset \cite{ref42} which combines the SNLS and SDSS SNe to create an extended sample of 740 SNe to reduce the estimation time, but we found that using the Pantheon data slightly improves the parameter estimations. In this case the chisquare is defined as 
\begin{equation}
\label{eq22} \chi^{2}_{SN}=(\mu(z)-\mu_{0})^{T}{\bf C}^{-1}(\mu(z)-\mu_{0}),
\end{equation}
where $\mu(z)$ is the predicted distance modulus given by
\begin{equation}
\label{eq23} \mu(z)=5\log_{10}[3000D_{c}(z)(1+z)]+25-5\log_{10}(h),
\end{equation}
${\bf C}^{-1}$ is the inverse of the $1048$ by $1048$ Pantheon compilation covariance matrix (${\bf C_{ij}}=\mbox{diag}(\sigma_{i}^{2})$), and $D_{c}(z)$ is the transverse co-moving distance defined as \cite{ref43}

\begin{equation}
\label{eq24} D_{c}(z)=|\Omega_{k}|^{-\frac{1}{2}}Sinn\left[|\Omega_{k}|^{\frac{1}{2}}\int_{0}^{z}\frac{dz'}{h(z')}\right],
\end{equation}
where $Sinn(x)= \sin(x), x, \sinh(x)$ for $\Omega_{k}<0, \Omega_{k}=0, \Omega_{k}>0$ respectively. It is worth mentioning that since the parameter $h (H_{0})$ is only an additive constant, thus, marginalizing over $h$ does not affect the
SNe results.\\

{\bf Observational Hubble Data}: we use OHD data from Table 2 of Ref \cite{ref33} which includes $31H(z)$ datapoints in the redshift range $.07\leq z\leq 1.965$. This measurements are uncorrelated and determined using the cosmic chronometric (CC) technique. It is worth nothing that the OHD data can be categorized into the following two classes, (1) BAO based data and (2) cosmic chronometric (CC) based data. To obtain OHD data from BAO, we usually model the redshift space distortions and assume an acoustic scale in a specific model. Therefore, this class of data is model-dependent and hence cannot be used for constraining a cosmological model. Nonetheless, to determine the CC data we use the most massive and passively evolving galaxies based on the “galaxy differential age ” method. consequently, this class of OHD data is model-independent (see ref\cite{ref44} for more details). Since this compilation includes uncorrelated data, we have ${\bf C_{ij}}=\mbox{diag}(\sigma_{i}^{2})$ as covariance matrix for this class of data. For OHD data the chisqure is given by
\begin{equation}
\label{eq25} \chi^{2}_{HOD}= (H(z)-H_{0})^{T}{\bf C}^{-1}(H(z)-H_{0}).
\end{equation}

{\bf BAO Data}: to obtain the BAO constraints on the model parameters we use ten numbers of $r_{s} (z_{d} )/D_{V} (z)$
extracted from the 6dFGS \cite{ref45}, SDSS-MGS \cite{ref46}, BOSS \cite{ref47}, BOSS CMASS \cite{ref48}, and WiggleZ \cite{ref49} surveys. Here, $D_{V}(z)$ is the effective distance measure related to the BAO scale define by
\begin{equation}
\label{eq26} D_{V}(z)=\left[r^{2}(z)\frac{cz}{H(z)}\right]^\frac{1}{3},
\end{equation}
where $c$ is the speed of light, $r_{s} (z_{d} )$ is the comoving sound horizon size at the drag epoch and $z_{d}$ is the redshift at which baryons are released from photons (see Ref \cite{ref35} for more details about these two parameters). Following to De Felice et al \cite{ref35}, the chisqure for BAO with such data could be written as

\begin{multline}
\chi^{2}_{BAO}=\\\frac{1}{0.015^{2}}\left[\frac{r_{s}(z_{d})}{D_{V}(z=0.106)}-0.336\right]^{2}
+\frac{1}{(\frac{25}{149.69})^{2}}\left[\frac{D_{V}(z=0.15)}{r_{s}(z_{d})}-\frac{664}{148.69}\right]^{2} 
+\frac{1}{(\frac{25}{149.28})^{2}}\left[\frac{D_{V}(z=0.32)}{r_{s}(z_{d})}-\frac{1264}{149.28}\right]^{2}\\
+\frac{1}{(\frac{16}{147.78})^{2}}\left[\frac{D_{V}(z=0.38)}{r_{s}(z_{d})}-\frac{1477}{147.78}\right]^{2}
+\frac{1}{0.0071^{2}}\left[\frac{r_{s}(z_{d})}{D_{V}(z=0.44)}-0.0916\right]^{2}
+\frac{1}{(\frac{19}{147.78})^{2}}\left[\frac{D_{V}(z=0.51)}{r_{s}(z_{d})}-\frac{1877}{147.78}\right]^{2}\\
+\frac{1}{(\frac{20}{149.28})^{2}}\left[\frac{D_{V}(z=0.57)}{r_{s}(z_{d})}-\frac{2056}{149.28}\right]^{2}
+\frac{1}{0.0034^{2}}\left[\frac{r_{s}(z_{d})}{D_{V}(z=0.6)}-0.0726\right]^{2}
+\frac{1}{(\frac{22}{147.78})^{2}}\left[\frac{D_{V}(z=0.61)}{r_{s}(z_{d})}-\frac{2140}{147.78}\right]^{2}\\
+\frac{1}{0.0032^{2}}\left[\frac{r_{s}(z_{d})}{D_{V}(z=0.73)}-0.0592\right]^{2}
\end{multline}
\\

Finally, since these three datasets are independent, the total chisqure could be written as $\chi^{2}_{tot}=\chi^{2}_{OHD}+\chi^{2}_{BAO}+\chi^{2}_{SN}$. Therefore, we evaluated the following total likelihood
\begin{equation}
\label{eq28} \mathcal{L}_{tot}\propto \exp\left(-\frac{1}{2} \chi^{2}_{tot}\right).
\end{equation}
Moreover, to check the degeneracy direction between computed parameters we perform covariance matrix which could be obtained from our MCMC runs. Theoretically, covariance matrix is defined as 
\begin{equation}
\label{eq29} C_{\alpha\beta}=\rho_{\alpha\beta}\sigma({\theta_{\alpha}})\sigma({\theta_{\beta}}),
\end{equation}
where the uncertainties in parameters $\theta_{\alpha}$ and $\theta_{\beta}$ are given by $\sigma({\theta_{\alpha}})$ and $\sigma({\theta_{\beta}})$ are the $1\sigma$ respectively, and $\rho_{\alpha\beta}$ is the correlation coefficient between $\theta_{\alpha}$ and $\theta_{\beta}$.\\

We use Metropolis-Hasting algorithm to generate MCMC chains for all parameters. For each parameter we run 4 parallel chains with 6000 separate iterations to stabilize the estimations. We perform Gelman-Rubin and Geweke tests to confirm the convergence of MCMC chains. We also confirm the convergence of all chains by monitoring the trace plots for good mixing and stationarity of the posterior distributions.
In our Bayesian estimations, we assume the following uniform priors for free parameters of the model (\ref{eq21}):
\begin{equation}
\label{eq30} H_{0} \sim U(50-100)~~~A \sim U(0-5)~~~j \sim U(-0.125-10).
\end{equation}
It is worth noting that we use eqs (\ref{eq10a}) - (\ref{eq10d}) to estimate other parameters of the space $\Theta$ as derived parameters in our mcmc code.
\section{Results}
\label{sec:5}
In Table.~\ref{tab:1} we have listed our statistical analysis on parameter space (\ref{eq21}) using OHD, BAO, SNIa, and their joint combination dataset at 1$\sigma$ error. It is worth mentioning that, although, SNIa data by itself is not sensitive to the universe expansion rate $H_{0}$, but in the joint analysis, this data constrains other parameters of the model which in turn affect the computation of $H_{0}$. That is why in Table.~\ref{tab:1} we observe a change in the value of $H_{0}$ when fitting model to the joint OHD+BAO+SNIa data. In Table~.(\ref{tab:2}) we have compered our obtained $H_{0}$ for OHD and OHD+BAO+SNIa data to those obtained by other researchers. Results of this table clearly show that when we use joint dataset our computed $H_{0}$ is in high agreement with those obtained by Sievers et al ($70\pm 2.4$) \cite{ref51}, Chen et al ($68.4^{+2.9}_{-3.3}$) \cite{ref52}, and J. Dunkley et al ($69.7\pm 2.5$) \cite{ref53}. However, while the estimated value of $H_{0}$ obtained from fitting to OHD data is in excellent agreement with those of Chen \& Ratra ($68\pm 2.8$) \cite{ref50} and Chen et al ($68.4^{+2.9}_{-3.3}$) \cite{ref52}, it is in high tension with what reported by Riess et al \cite{ref58}. In all cases, our estimated $H_{0}$ is in good agreement with results obtained using non-parametric approaches \cite{ref31,ref32}.\\

We have also compared our estimated deceleration parameter to those of Aviles et al (columns forth and fifth of Table 1 \cite{ref18}), Muthukrishna \& Parkinson (row third of Table 3 \cite{ref22}), zhang et al (row forth of Table 1 \cite{ref20}) and G\'{o}mez-Valent \cite{ref32} in Table~.(\ref{tab:3}).
\begin{table*}[h!]
\caption{Best fit value and 1$\sigma$ error bars for each cosmographic parameters. We perform a fit by using H(z) only (column two), BAO data only (column three), SNIa data only (column four), and by using the combined H(z), BAO and SNIa data together (column five).}
\centering
\setlength{\tabcolsep}{5pt}
\scalebox{0.9}{
\begin{tabular} {cccccc}
\hline
Parameter    & OHD (CC) & BAO & SNIa (Pantheon) & CC + BAO + Pantheon \\[0.5ex]           
\hline
\hline{\smallskip}
$H_{0}$ & $67.9\pm 2.9$ & $-$ & $-$ & $69.9\pm 1.7$\\[0.2cm]              
		
$A$ &  $0.328^{+0.052}_{-0.063}$  & $0.24\pm 0.14$ & $0.278^{+0.014}_{-0.016}$ &  $0.279^{+0.013}_{-0.017}$ &\\[0.2cm]        
		
$B$ &  $0.672^{+0.063}_{-0.052}$ & $0.76\pm 0.14$ & $0.722^{+0.016}_{-0.014}$ & $0.721^{+0.017}_{-0.013}$\\[0.2cm]  
		
$q_{0}$ & $-0.508^{+0.080}_{-0.096}$ & $-0.64\pm 0.21$  & $-0.590^{+0.024}_{-0.030}$ & $-0.587^{+0.023}_{-0.032}$\\[0.2cm]

$\Omega_{0k}$ & $0.001\pm 0.058$ & $0.001\pm0.057$ & $-0.039^{+0.019}_{-0.059}$ & $-0.038^{+0.023}_{-0.061}$\\[0.2cm]
		
$j_{0}$ & $0.999\pm 0.057 $ & $0.999\pm 0.057$ & $1.039^{+0.059}_{-0.019}$ & $1.038^{+0.061}_{-0.023}$\\[0.2cm]
		
$s_{0}$ & $-0.47^{+0.28}_{-0.23}$ & $-0.09\pm 0.65$ & $-0.299\pm 0.059$ & $-0.308\pm 0.058$\\[0.2cm]
		
$l_{0}$ & $-2.55^{+0.83}_{-1.2}$ & $-3.7^{+1.9}_{-3.2}$ & $-3.35\pm 0.21$ & $-3.32\pm 0.21$\\[0.5ex]

\hline
\end{tabular}}
\label{tab:1}
\end{table*}
\begin{figure*}[h!]
\centering
\includegraphics[width=18cm,height=18cm,angle=0]{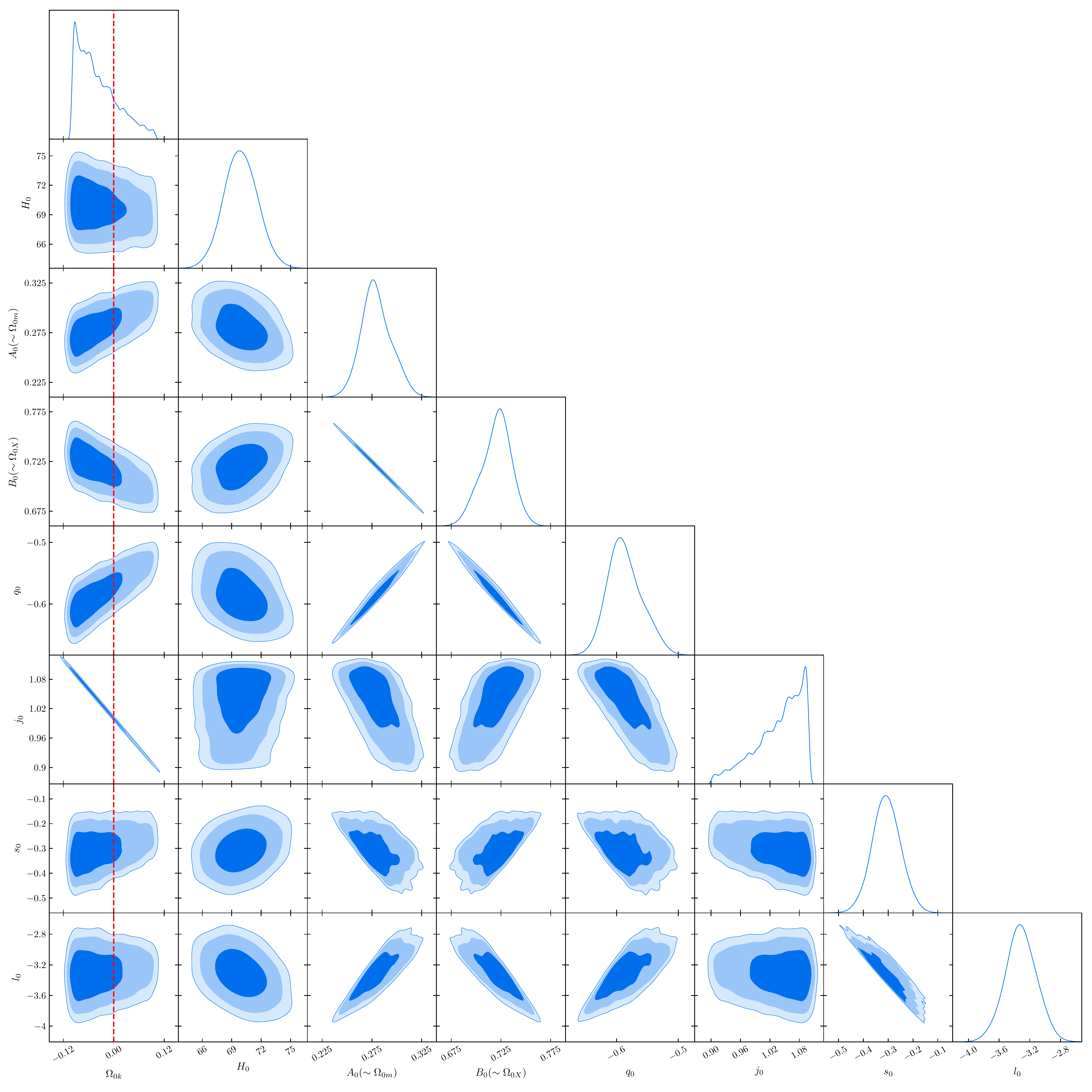}
\caption{One-dimensional marginalized distribution, and three-dimensional contours with $68\%$ CL, $95\%$ CL, and $99\%$ CL for some parameters from parameter space {\bf$\Theta$} using {\bf CC+BAO+Pantheon} data. The vertical dashed red line stands for $\Omega_{0k}=0$.}
\label{fig1}
\end{figure*}
It is worth noting that while the estimated values of $j_{0}, l_{0}$ reported in Ref \cite{ref22} almost not physical (taking high values), our computed values of these parameters are physical for  OHD, BAO, SNIa and their joint dataset. When we fit {\bf$\Theta$} to OHD data alone we obtain $A(\sim\Omega_{m})=0.328^{+0.052}_{-0.063}$ and $B(\sim\Omega_{X})=0.672^{+0.063}_{-0.052}$, these results are in excellent agreement with those obtained from 9years WMAP \cite{ref57}. Also when we fit {\bf$\Theta$} to SNIa data alone, we obtain $A(\sim\Omega_{m})=0.278^{+0.014}_{-0.016}(0.279^{+0.014}_{-0.016}) $ and $B(\sim\Omega_{X})=0.722^{+0.016}_{-0.014}(0.721^{+0.016}_{-0.014})$, these results are in excellent agreement with those obtained from Planck (2015) collaboration \cite{ref58}. The estimated values of $A$ \& $B$ obtained from fitting model over BAO data are also in good agreement with those of ref \cite{ref58}. Recently Mukherjee \&  Banerjee \cite{ref41} have used OHD, SNIa, and BAO data to constrain the same model given by eq \ref{eq8} and obtained $A=0.286\pm0.015$ \& $j=-1.027\pm0.037$ (see Table 1 of this reference). Comparing our estimated values for $A$ and $j$ reported in Table.~\ref{tab:1} with those obtained in ref \cite{ref41} shows a very high agreement.
\begin{table}[h]
\caption{The value of $H_{0}$ obtained by different researches.}
\centering
\setlength{\tabcolsep}{5pt}
\scalebox{0.7}{
\begin{tabular} {cccc}
\hline
Researchers     &   $H_{0}$  &  Reference \\[0.5ex]
\hline
\hline{\smallskip}
G\'{o}mez-Valent \& Amendola    & $68.9\pm 1.96$ (at $68\%$)   & \cite{ref31}    \\
			
G\'{o}mez-Valent    & $70.45\pm 2.36$ (at $68\%$)   & \cite{ref32}    \\
			
Ade et al (Planck 2015)    & $67.8\pm 0.9$ (at $68\%$)      & \cite{ref54}    \\
			
Chen \& Ratra              & $68\pm 2.8$  (at $68\%$) &  \cite{ref50}   \\
			
Sievers et al              & $70\pm 2.4$ (at $68\%$)     & \cite{ref51}    \\
			
Gott et al            & $67\pm 3.5$  (at $68\%$)    & \cite{ref55}     \\
			
J. Dunkley et al (CMB)     & $69.7\pm 2.5$   (at $68\%$)   & \cite{ref53}    \\
			
Aubourg et al (BAO)        & $67.3 \pm 1.1 $  (at $68\%$)    & \cite{ref56} \\
			
V. Lukovic et al            & $66.5\pm 1.8$  (at $68\%$)   & \cite{ref57}      \\
			
Chen et al            & $68.4^{+2.9}_{-3.3}$  (at $68\%$)   & \cite{ref52}       \\
			
Riess et al           & $73.24\pm1.74$  (at $68\%$)   & \cite{ref58}     \\ 
			
Present work          & $67.9\pm 2.9$ (at $68\%$)&   for OHD\\
	
Present work          & $69.9\pm 1.7$ (at $68\%$)&  for CC+BAO+Pantheon\\ [0.5ex]
\hline
\end{tabular}}
\label{tab:2}
\end{table}

We have compared the computed spatial curvature $\Omega_{k}$ to those of 9 years WMAP \cite{ref59}, Planck (2015) collaboration and Park\& Ratra \cite{ref60} in Table.~\ref{tab:4}. From this table we observe that while individual OHD (CC) or BAO data predict (estimate) a slightly open universe ($\Omega_{0k}>0$), SNIa (Pantheon) and OHD+BAO+SNIa predict a slightly closed universe ($\Omega_{0k}<0$). Moreover, from this table, we see that OHD (BAO) data alone put  tighter constrain on this parameter. It is worth nothing a stringent test of eternal inflation could be provided by constraining $\Omega_{k}$ at around the $10^{-4}$ level \cite{ref61,ref62,ref63}. In fact, bellow $\sim 10^{-4}$, $\Omega_{k}$ cannot be decisively distinguished from primordial fluctuations \cite{ref64}. However, large-scale anomalies as well as some inflationary scenarios tend to have observable levels of spatial curvature \cite{ref65,ref66}.\\

\begin{figure*}[h!]
\begin{minipage}[b]{0.5\linewidth}
\centering
\includegraphics[width=8cm,height=6cm,angle=0]{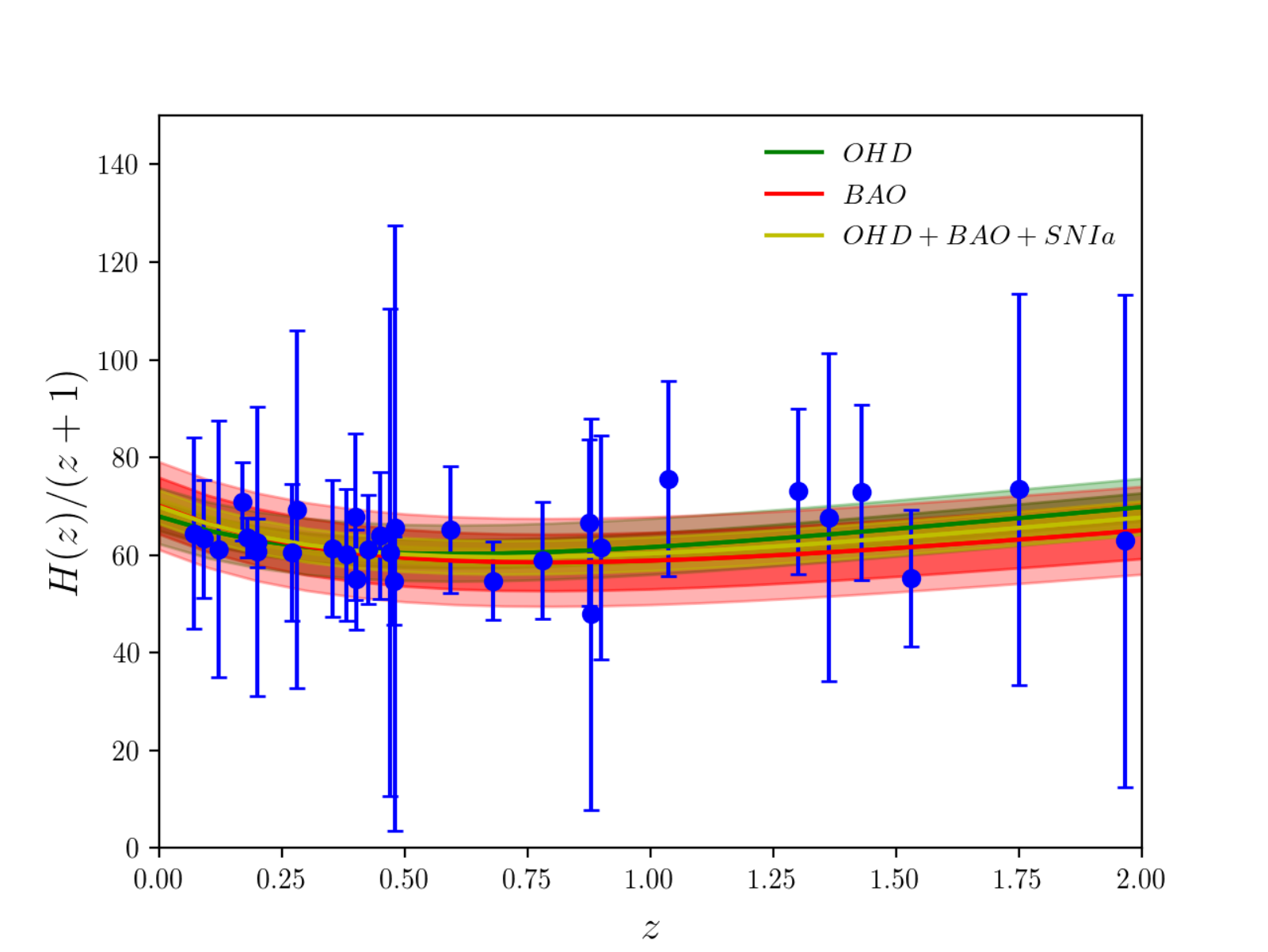} \\
\caption{The plot of Hubble rate  versus the redshift $z$ at 1$\sigma$ and 2$\sigma$ confidence level for OHD (green color), BAO (red color) and OHD+BAO+SNIa (yellow color). The points with bars indicate the experimental data summarized in Table 2 of Ref \cite{ref33}. It is clear that using joint datasets gives raise to better fit to the data.}
\label{fig2}
\end{minipage}
\hspace{0.5cm}
\begin{minipage}[b]{0.5\linewidth}
\centering
\includegraphics[width=8cm,height=6cm,angle=0]{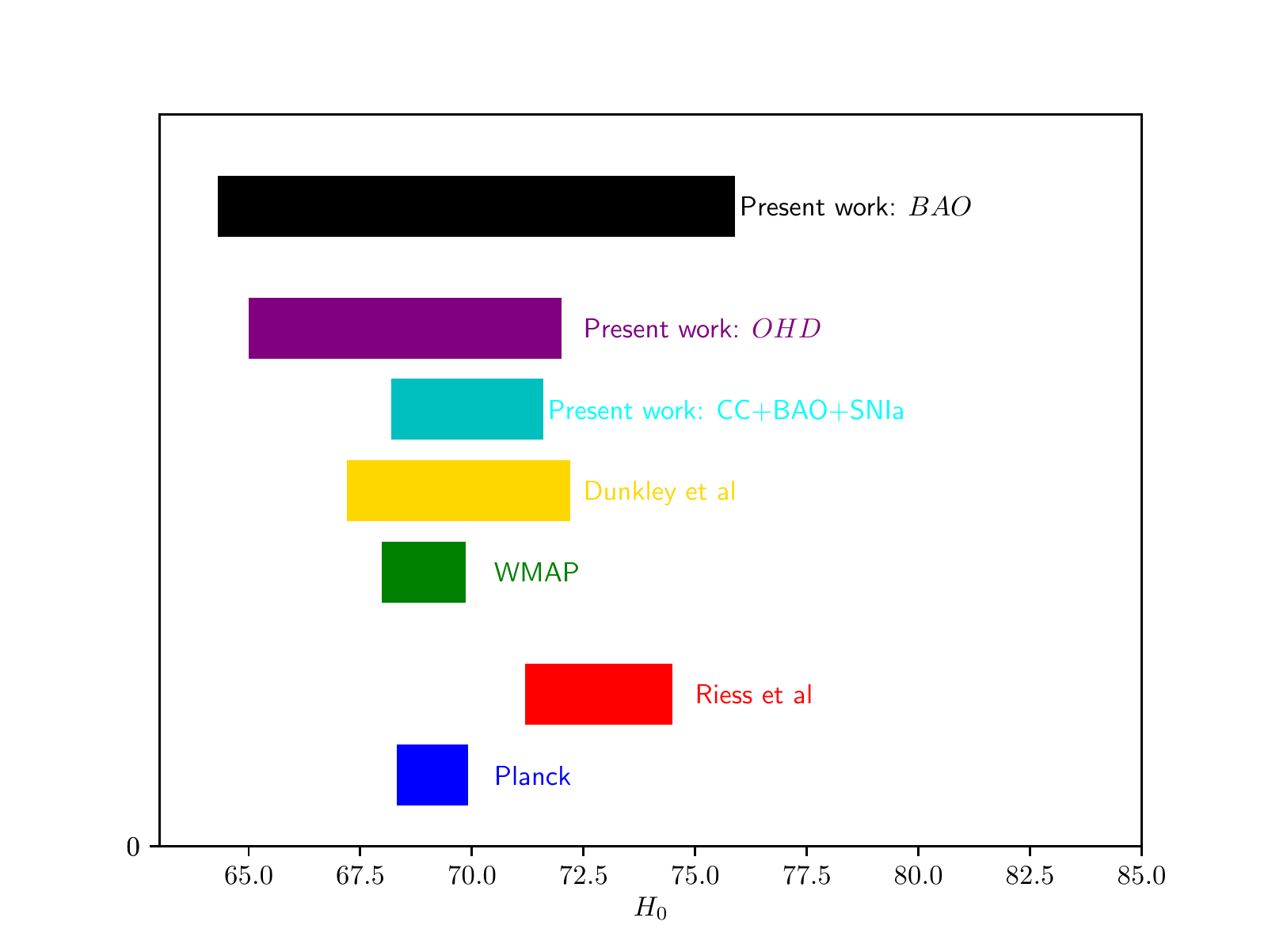}
\caption{Schematic representation of $H_{0}$ (at 1$\sigma$) for model \ref{eq16} ($OHD$(purple color), $BAO$ (black color) and $CC+BAO+SNIa$(cyan color)). Constraints from the direct measurement by Riess et al. (2016) (red color) WMAP (green color), Dunkley et al (gold color), and Planck (2015) (blue color) are also shown.}
\label{fig3}
\end{minipage}
\end{figure*}

\begin{figure*}[h!]
\centering
\begin{subfigure}[b]{0.23\textwidth}
\centering
\includegraphics[width=\textwidth]{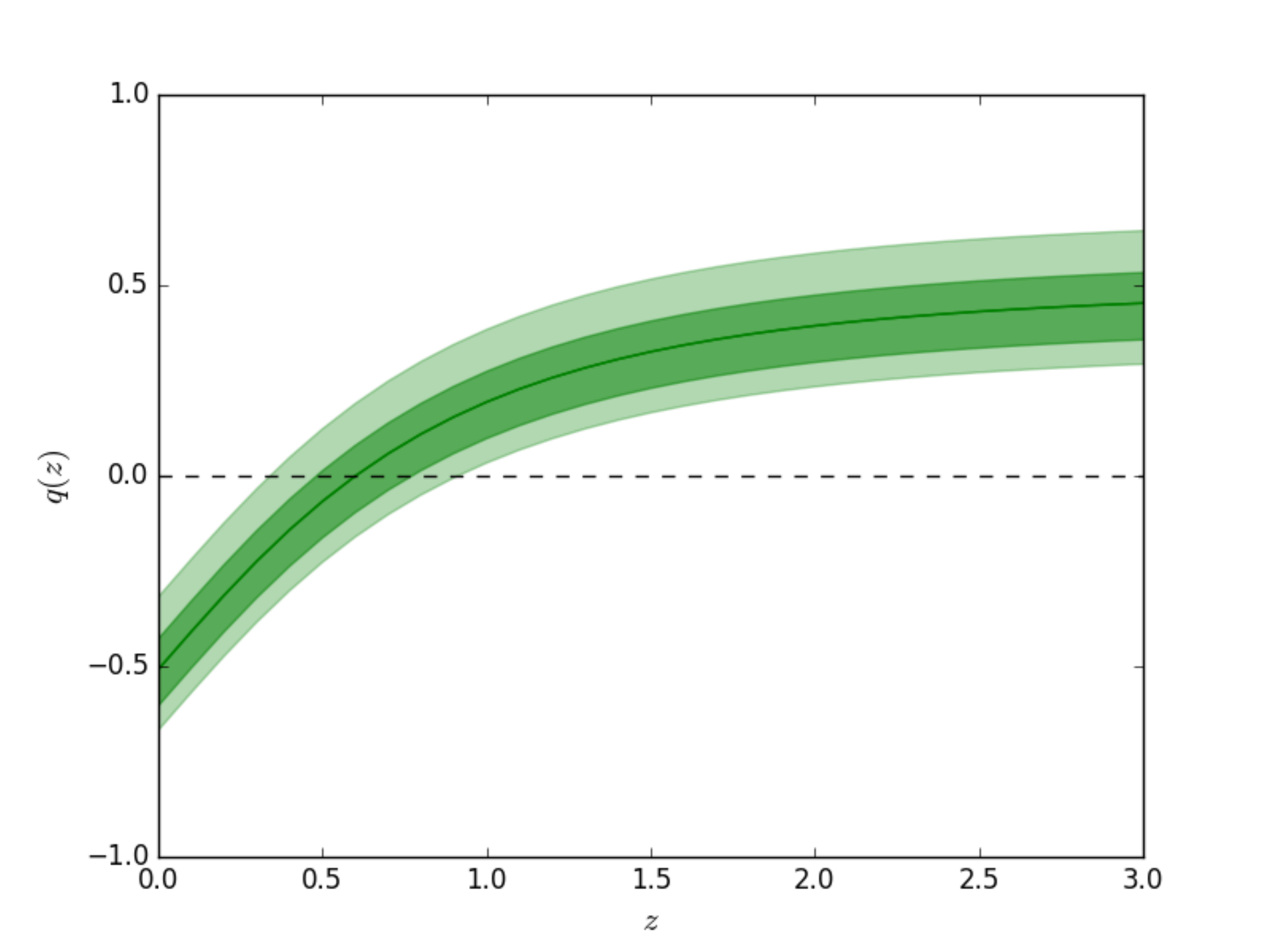}
\caption{OHD}
\label{fig4a}
\end{subfigure}%
\begin{subfigure}[b]{0.23\textwidth}
\centering
\includegraphics[width=\textwidth]{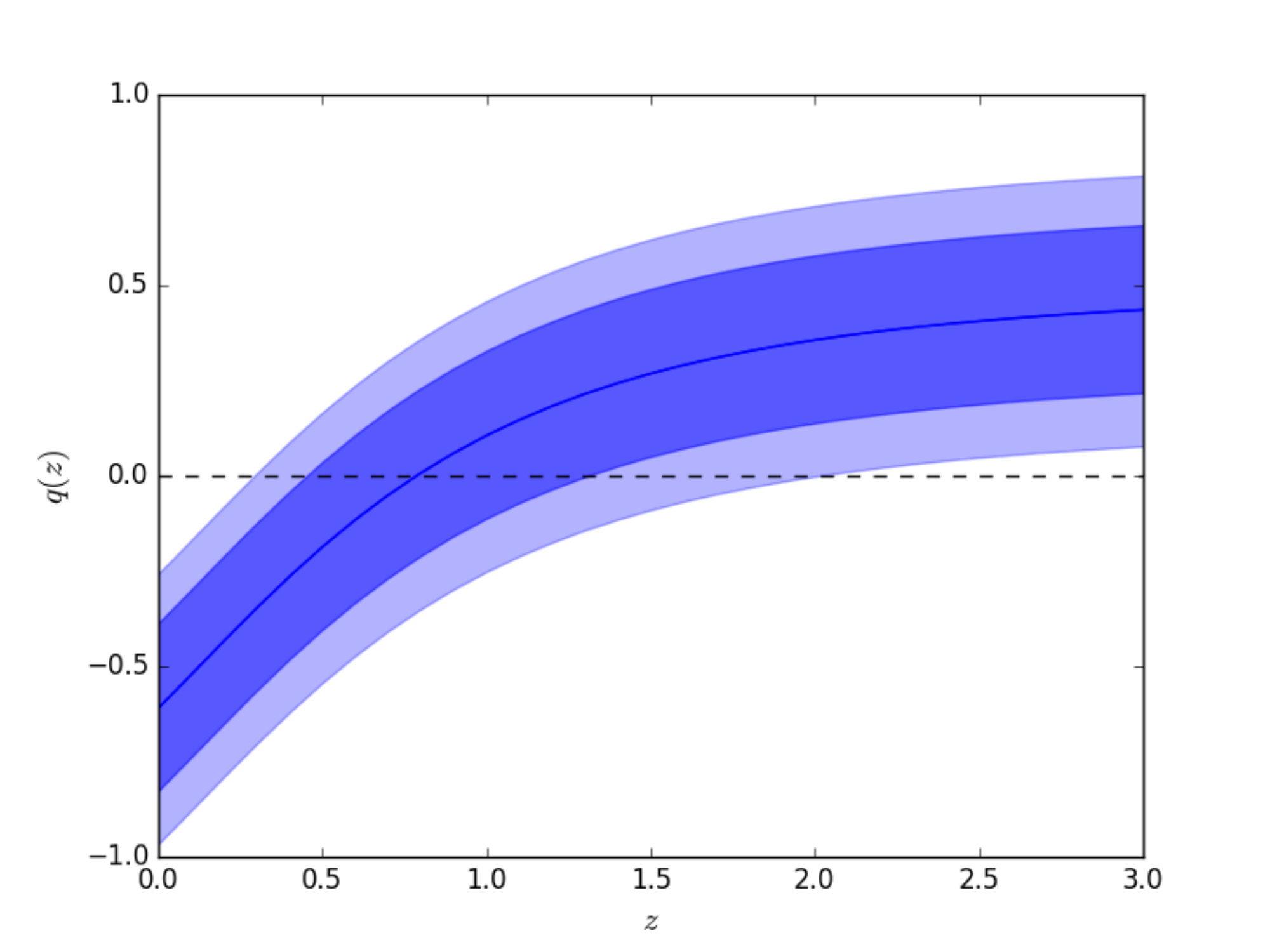}
\caption{BAO}
\label{fig4b}
\end{subfigure}
\begin{subfigure}[b]{0.23\textwidth}
		
\centering
\includegraphics[width=\textwidth]{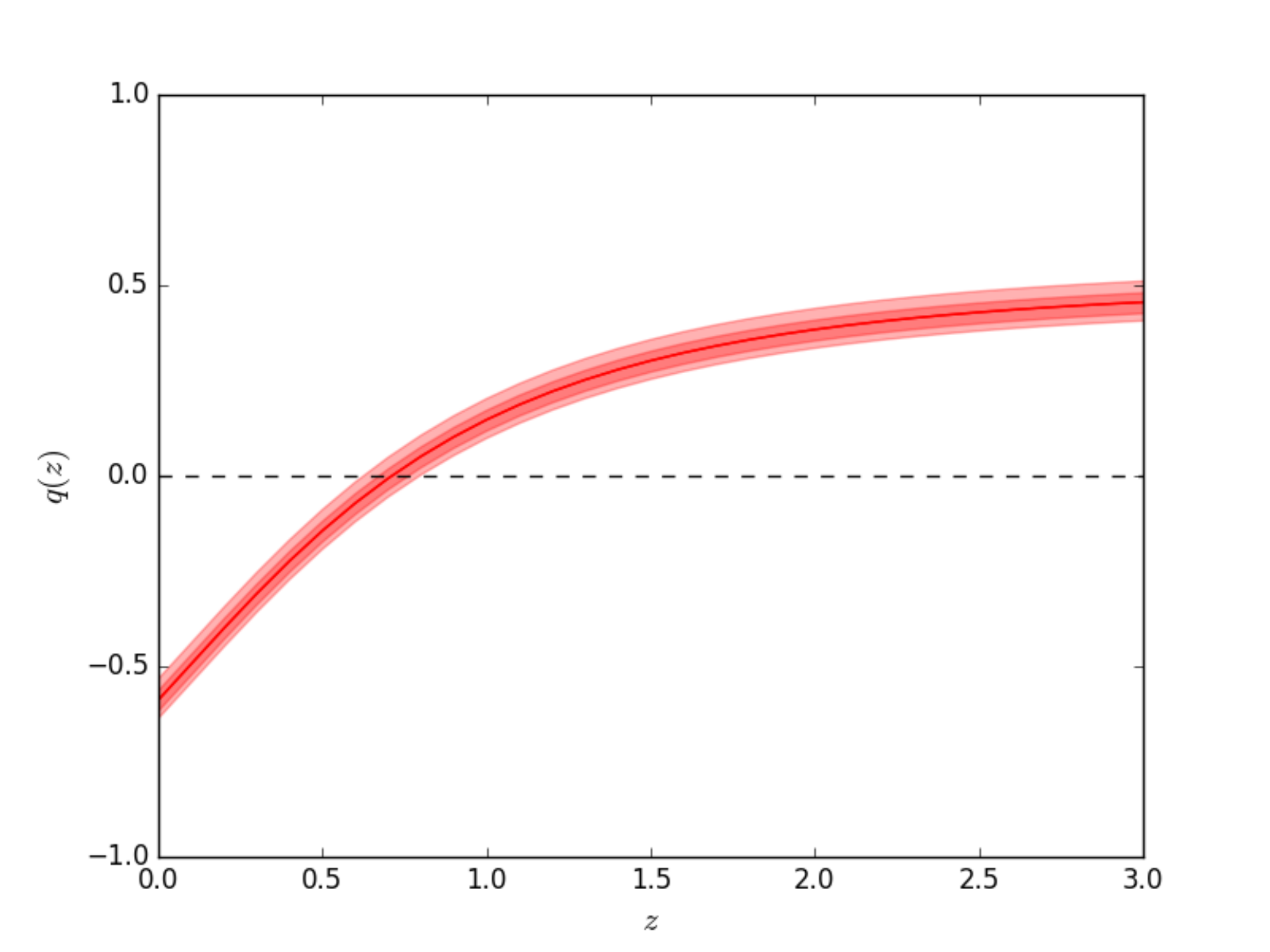}
\caption{SNIa}
\label{fig4c}
\end{subfigure}
\begin{subfigure}[b]{0.23\textwidth}
\centering
\includegraphics[width=\textwidth]{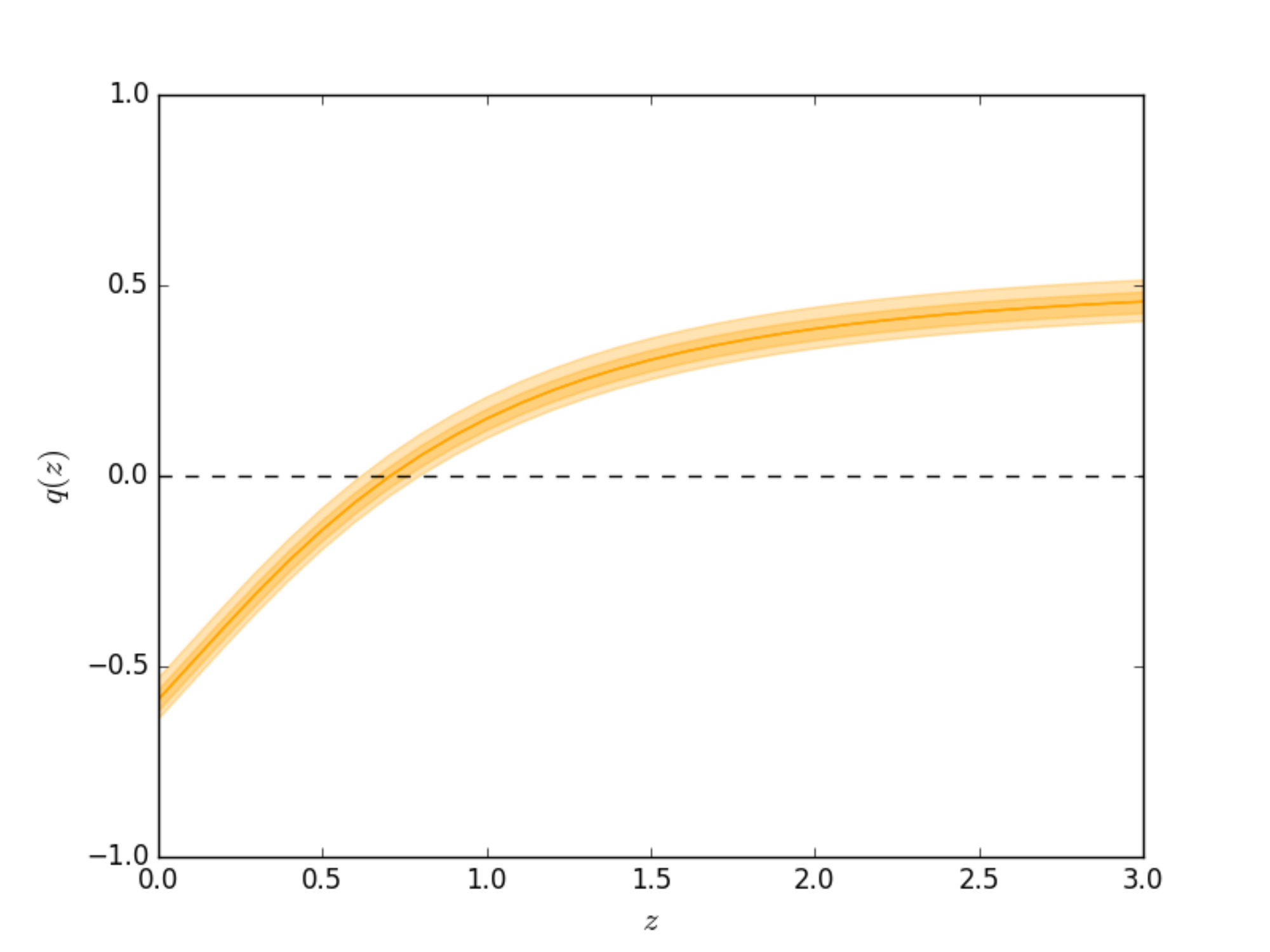}
\caption{OHD+BAO+SNIa}
\label{fig4d}
\end{subfigure}
\caption{The plots of deceleration parameters at 1$\sigma$ and 2$\sigma$ confidence level. Green, Blue, Red and orange figures show our fit to OHD, BAO, SNIa, and OHD+BAO+SNIa respectively.}
\label{fig4}
\end{figure*}

\begin{table*}[h!]
\caption{The value of $q_{0}$ at 1$\sigma$ obtained by different researches. Note that we have used Pantheon compilation.}
\centering
\scalebox{0.8}{
\begin{tabular} {ccccccc}
\hline
Researchers     &   OHD (CC) & BAO & SNIa& OHD+SNIa &CC+ABO+Pantheon &Reference \\[0.5ex]
\hline
\hline{\smallskip}
Aviles et al    & $-$&$-$ &$-0.6250^{+0.5580}_{-0.4953}$ & $-0.6361^{+0.3720}_{-0.3645}$ & $-$&\cite{ref18}\\
			
Zhang et al    & $-$ & $-$& $-0.58\pm 0.29$ & $-$ & $-$& \cite{ref20}  \\
			
Muthukrishna \& Parkinson  & $-$ &$-$ & $-0.63^{+0.83}_{-0.52}$ & $-$ & $-$& \cite{ref22}\\
			
G\'{o}mez-Valent   & $-$ & $-$ &$-$ & $-$& $-0.554\pm 0.32$ & \cite{ref32}    \\
			
Present work   & $-0.508^{+0.080}_{-0.096}$& $-0.62\pm 0.22$&$-0.575^{+0.11}_{-0.095}$ & $-0.586^{+0.024}_{-0.032}$  & $-0.587^{+0.024}_{-0.031}$\\
[0.5ex]
\hline
\end{tabular}}
\label{tab:3}
\end{table*} 
The contour plots, at $1\sigma$, $2\sigma$, and $3\sigma$ confidence levels, of the parameter space {\bf$\Theta$} for joint OHD+BAO+SNIa are depicted in Figure.~\ref{fig1}. We have also depicted the robustness of our fits for  $H(z)$ in Figures.~\ref{fig2},~\ref{fig3}. Figure.~\ref{fig4} shows the variations of $q(z)$ at 68\% and 95\% error for these data and their joint combination. From these figures it is also clear that our exact method gives raise to much better and tighter constrains on CS parameter with respect to the previous works.

\begin{figure*}[h!]
\centering
\begin{subfigure}[b]{0.23\textwidth}
\centering
\includegraphics[width=\textwidth]{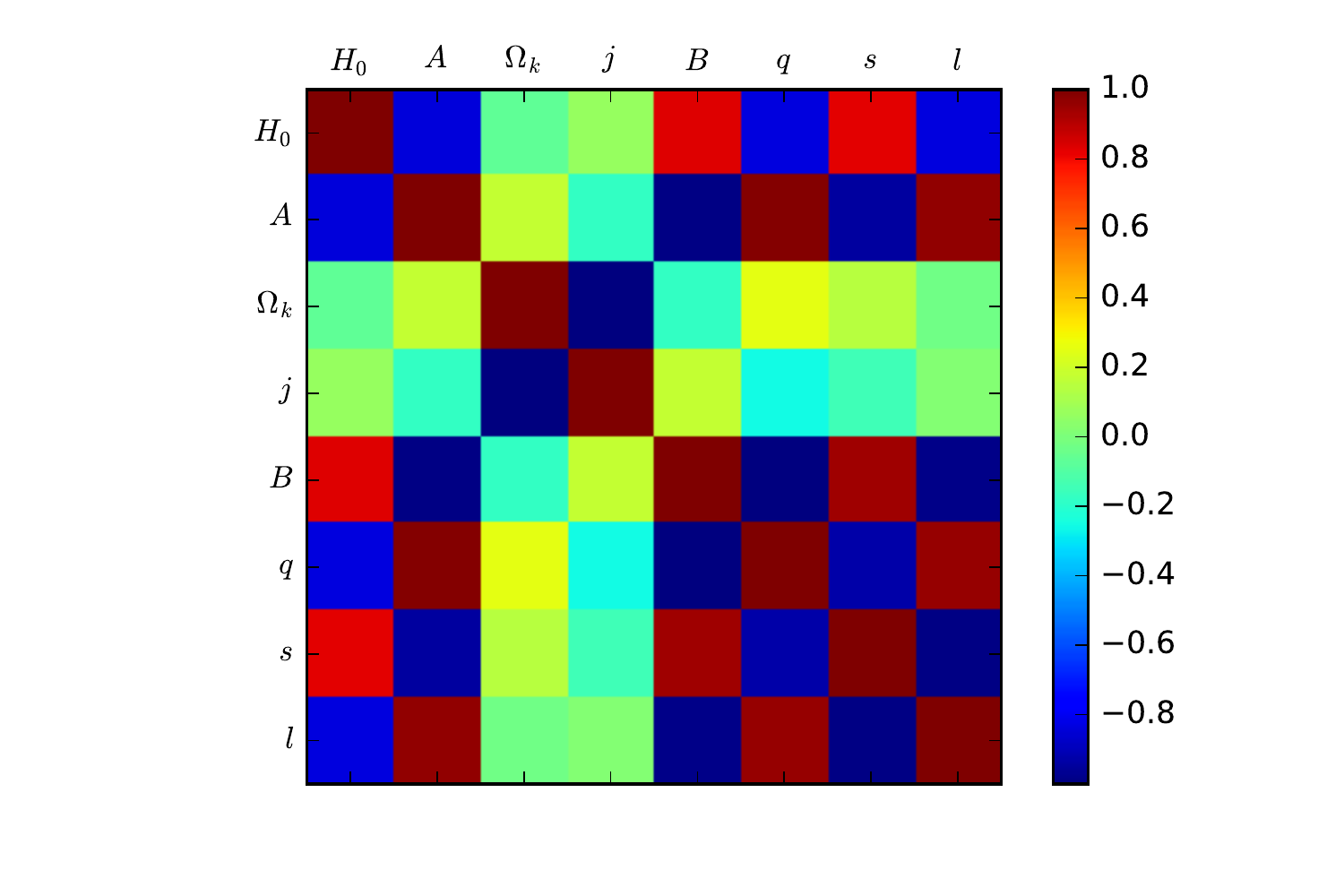}
\caption{OHD}
\label{fig5a}
\end{subfigure}%
\begin{subfigure}[b]{0.23\textwidth}
\centering
\includegraphics[width=\textwidth]{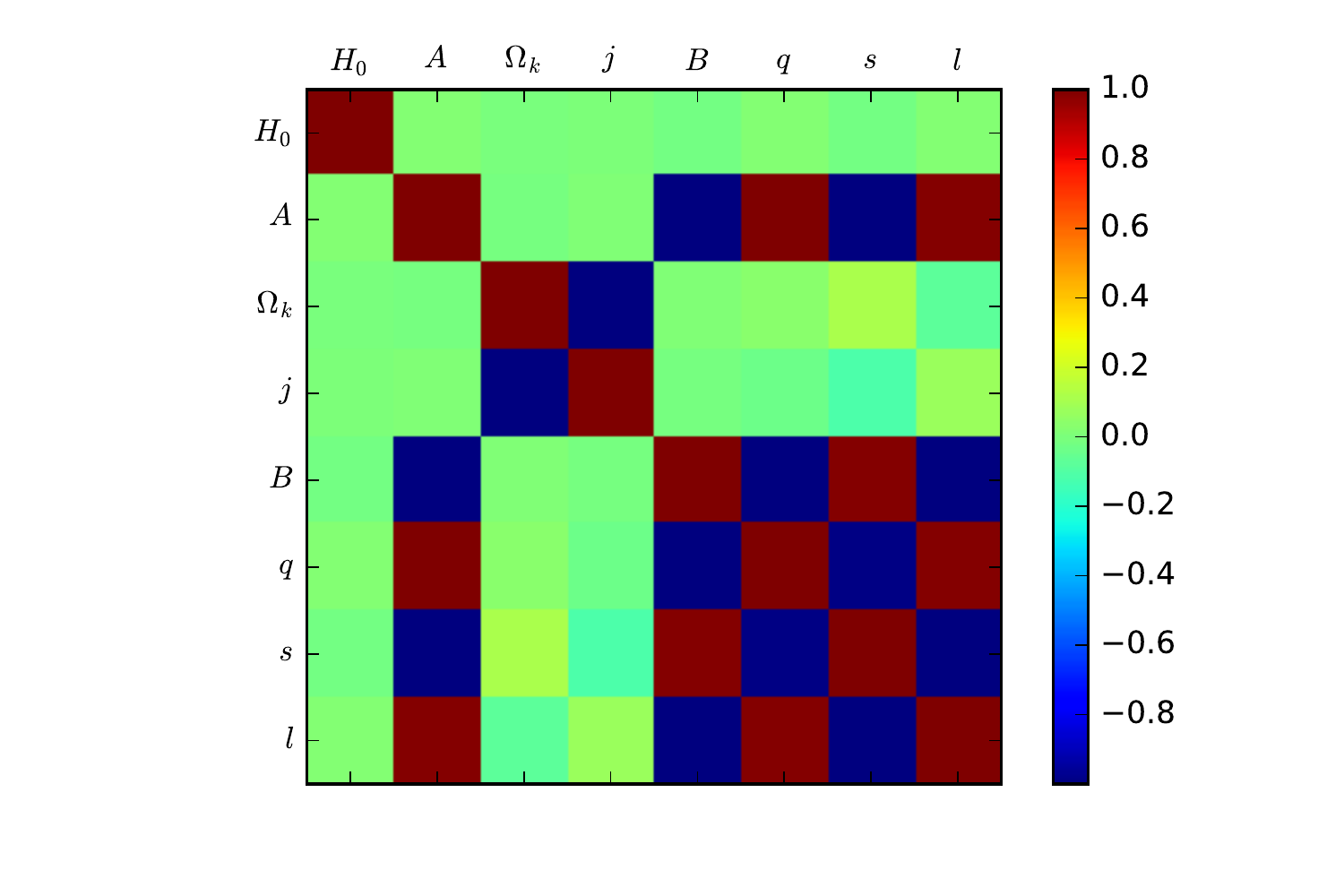}
\caption{BAO}
\label{fig5b}
\end{subfigure}
\begin{subfigure}[b]{0.23\textwidth}

\centering
\includegraphics[width=\textwidth]{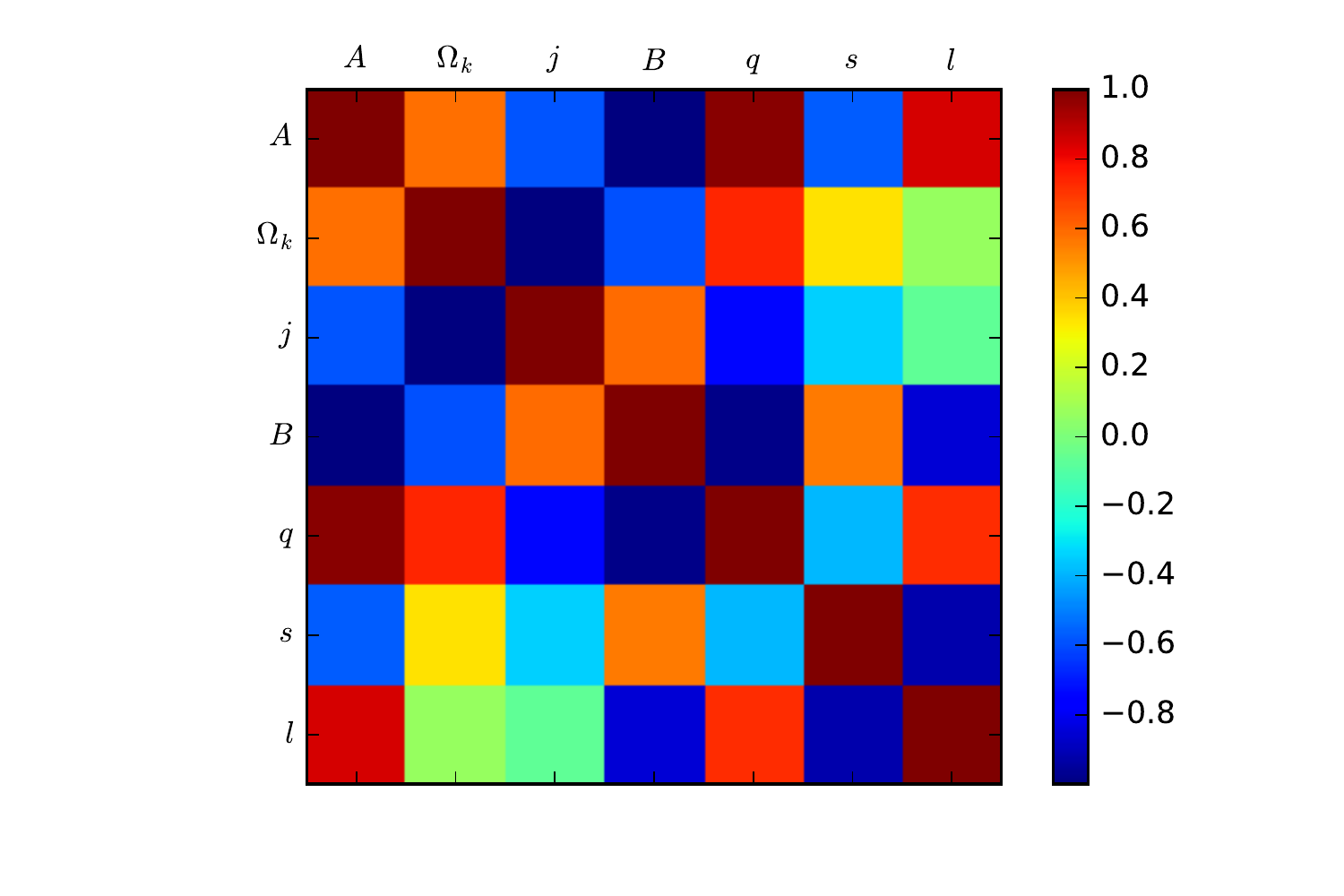}
\caption{SNIa}
\label{fig5c}
\end{subfigure}
\begin{subfigure}[b]{0.23\textwidth}
\centering
\includegraphics[width=\textwidth]{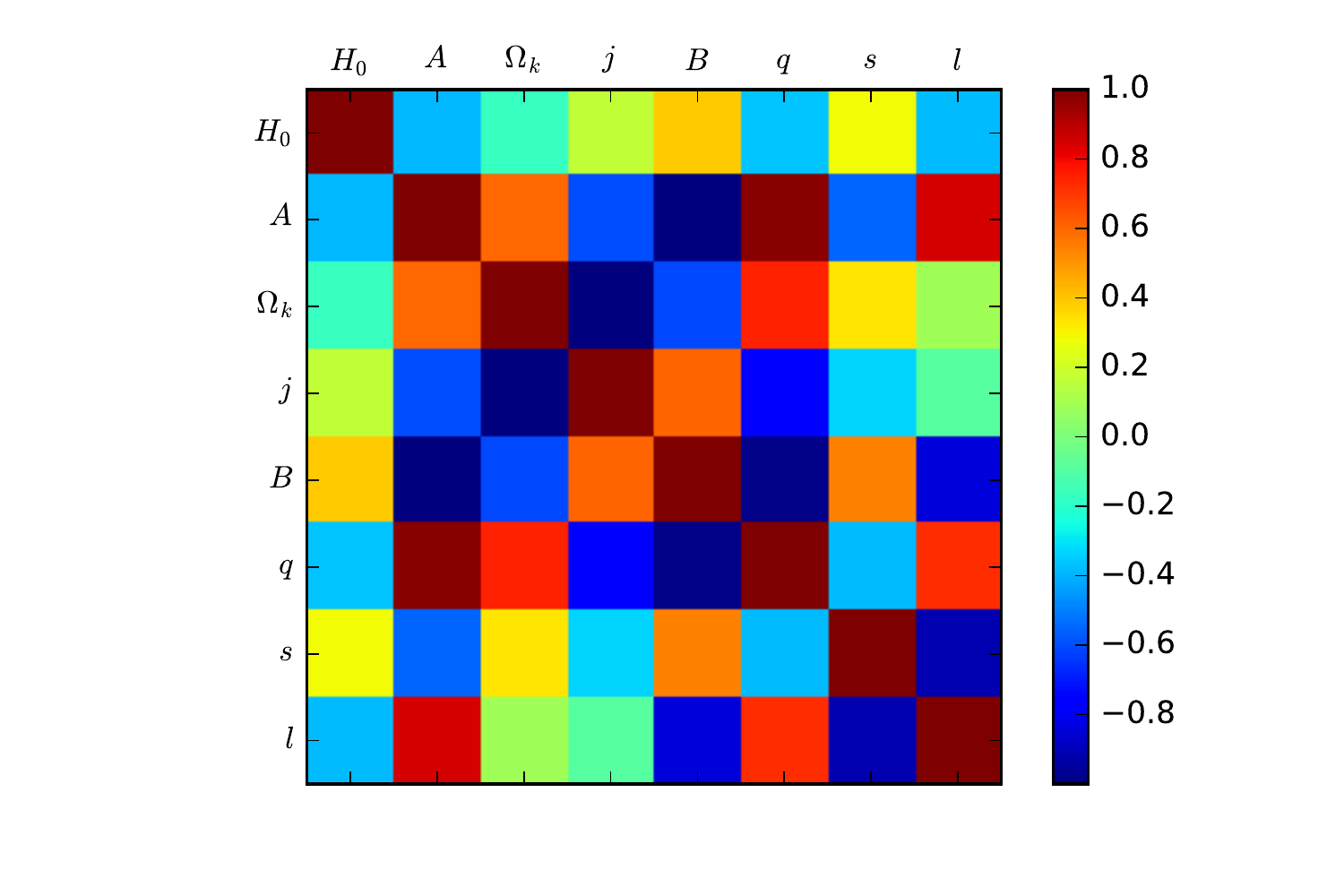}
\caption{OHD+BAO+SNIa}
\label{fig5d}
\end{subfigure}
\caption{Plots of correlation matrix of parameter space $\Theta$ using: (a) Hubble (OHD), (b) BAO, (c) SNIa, and (d) OHD+BAO+SNIa data. The color bars share the same scale.}
\label{fig5}
\end{figure*}

Figure.~\ref{fig5} depicts the correlation matrix for OHD (\ref{fig5a}), BAO (\ref{fig5b}), SNIa (\ref{fig5c}) and OHD+BAO+SNIa (\ref{fig5d}). 

\begin{table}[h!]
\caption{The value of $\Omega_{0k}$ at 1$\sigma$ obtained by different researches.}
\centering
\setlength{\tabcolsep}{2pt}
\scalebox{0.8}{
\begin{tabular} {ccccccc}
\hline
Researchers     & 9years WMAP & Planck (2015) & Park \& Ratra & This work\\[0.5ex]
\hline
\hline{\smallskip}
$\Omega_{0k}$    & $-0.0027^{+0.0039}_{-0.0038}$& $0.0008^{+0.0040}_{-0.0039} $ & $-0.0083\pm0.0016$& $-0.038^{+0.034}_{-0.061}$\\[0.5ex]
\hline
\end{tabular}}
\label{tab:4}
\end{table}
\subsection{Transition Redshift}
\label{subsec:5.1}
\begin{figure*}[ht]
\begin{minipage}[b]{0.5\linewidth}
\centering
\includegraphics[width=8cm,height=6cm,angle=0]{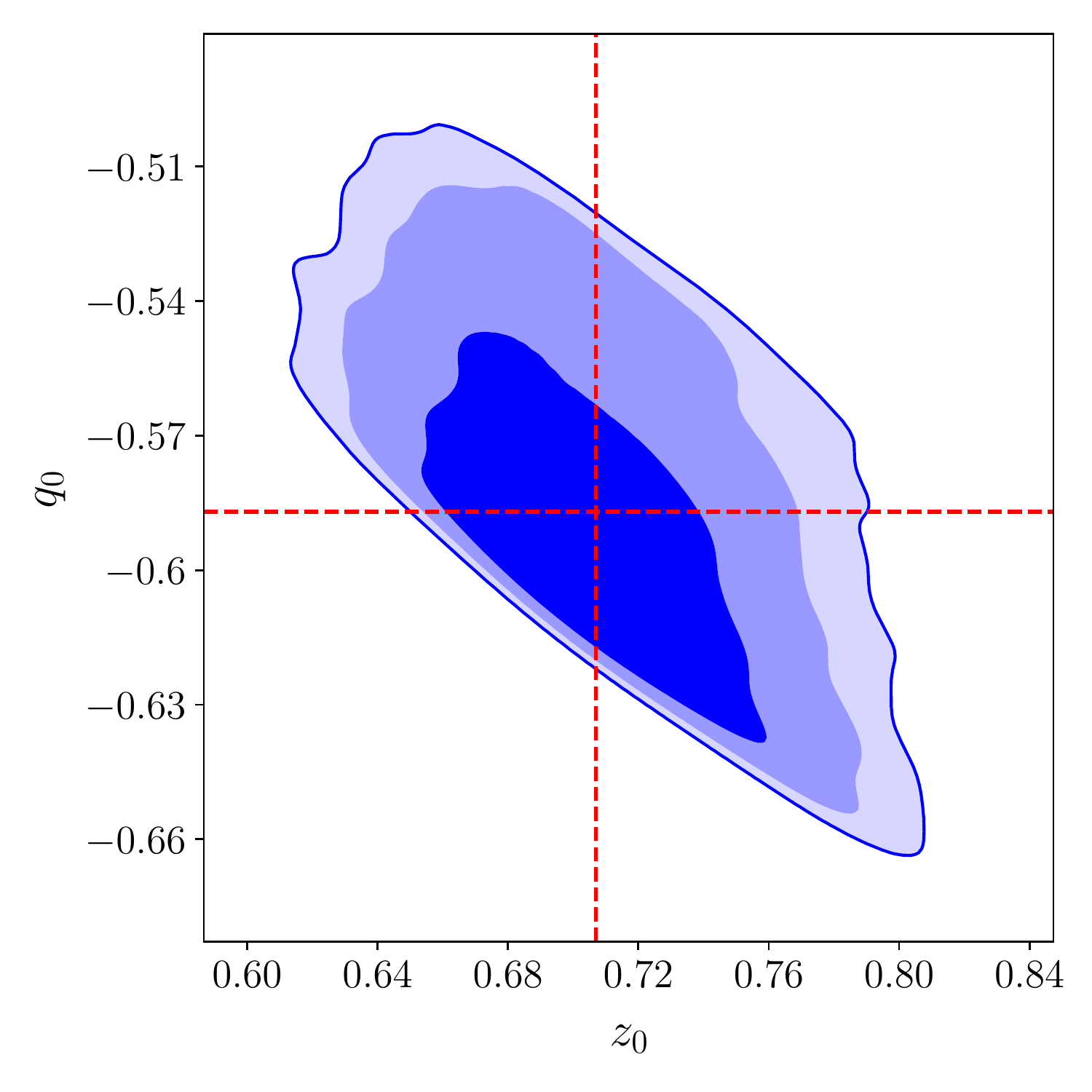} \\
\caption{Three-dimensional contours with $68\%$ CL, $95\%$ CL, and $99\%$ CL in  $q_{0}-z_{0t}$ plane using {\bf OHD+BAO+SNIa} data. The horizontal and vertical dashed lines stand for the best fit values.}
\label{fig6}
\end{minipage}
\hspace{0.5cm}
\begin{minipage}[b]{0.5\linewidth}
\centering
\includegraphics[width=8cm,height=6cm,angle=0]{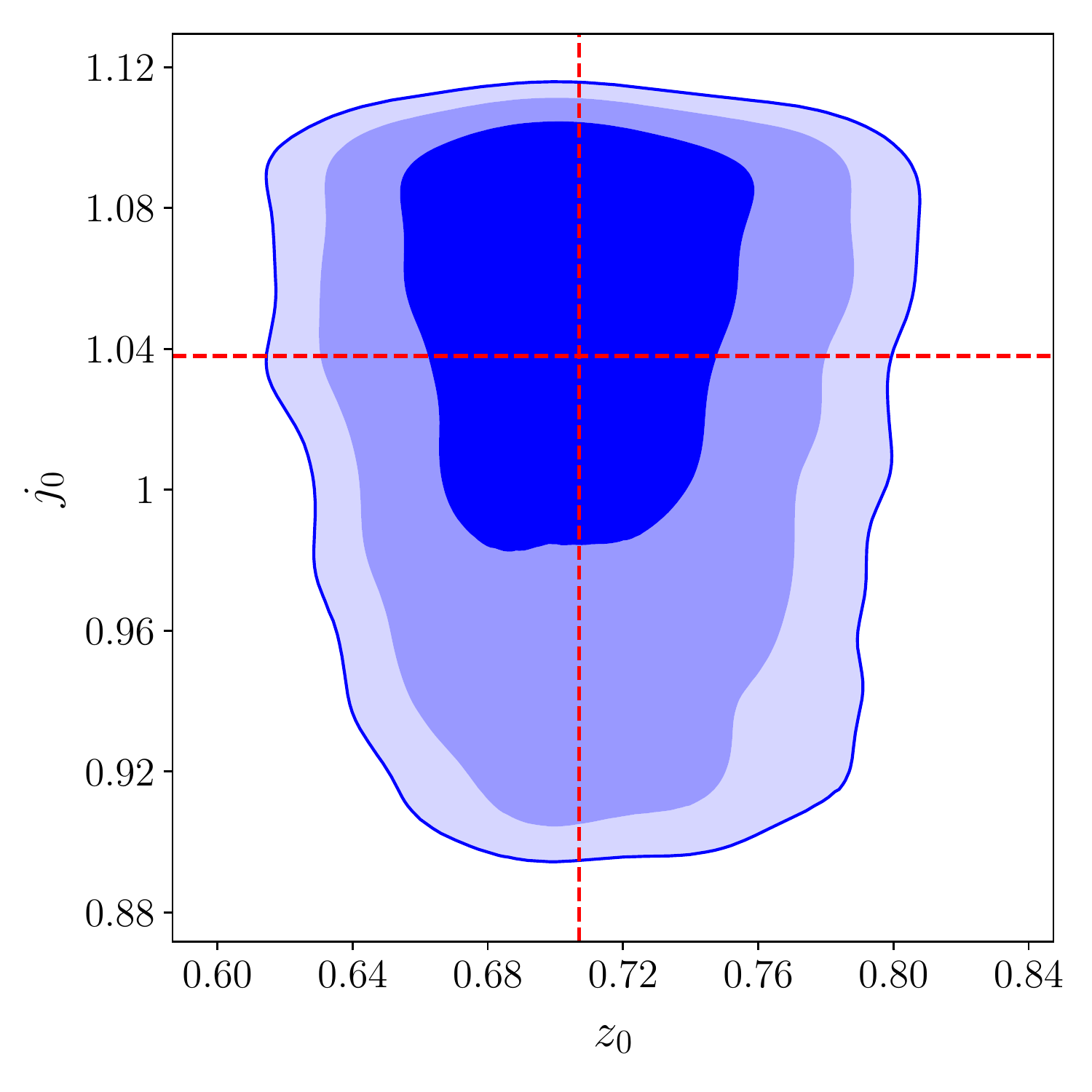}
\caption{Three-dimensional contours with $68\%$ CL, $95\%$ CL, and $99\%$ CL in  $j_{0}-z_{0t}$ plane using {\bf OHD+BAO+SNIa} data. The horizontal and vertical dashed lines stand for the best fit values.}
\label{fig7}
\end{minipage}
\end{figure*}

It is well known that the universe expansion phase has changed from decelerating to accelerating at a specific redshift called \textgravedbl Transition redshift $z_{t}$\textacutedbl \cite{ref67,ref68}. Mathematically we can find this decelerating-accelerating redshift by imposing $q(z)=\ddot{a}=0$ in (\ref{eq10a}). Doing so, and after some algebra we find the following general transition redshift.
\begin{equation}
\label{eq31} z_{t}=\left(-\dfrac{B(Y-2)}{A(X-2)}\right)^{\dfrac{1}{X-Y}}-1,
\end{equation}
which can be simplified and rewritten in terms of CS parameters as
\begin{equation}
\label{eq32} z_{t}=\left[-\dfrac{B(\sqrt{1+8j}+1)}{A(\sqrt{1+8j}-1)}\right]^{\dfrac{1}{\sqrt{1+8j}}}-1.
\end{equation}
Note that we may set $A=\sim\Omega_{m}$ and $B=\sim\Omega_{X}$ in above equation.  According to the previous discussion, for $\Lambda$CDM model, we can rewrite eq \ref{eq32} and relate transition redshift to the spacetime curvature as follows
\begin{equation}
\label{eq33} z_{t}=\left[-\dfrac{B(\sqrt{1+8j}+1)}{A(\sqrt{1+8j}-1)}\right]^{\dfrac{1}{\sqrt{9-8\Omega_{k}}}}-1.
\end{equation}
Our statistical analysis on transition redshift for both datasets and their joint combination could be seen in Table.~\ref{tab:5}. Figures.~\ref{fig6} and ~\ref{fig7} depict the robustness of our fits for $z_{t}$. Gastri et al \cite{ref69} have recently have used different SNIa data in combination with BAO data to constrain transition redshift. For MLCS2k2+BAO/CMB (see\cite{ref70} for details of MLCS2k2 data) they obtained $z_{t}=0.56^{+0.13}_{-0.10}$ and for SALT2+BAO/CMB (see \cite{ref71} for details of SALT2 data) they found $z_{t}=0.64^{+0.13}_{-0.07}$. Also Farooq et al \cite{ref72} used 38 $H(z)$ data and found $z_{t}=0.72\pm0.05(0.84\pm0.03)$ for two Hubble constant priors as $H_{0}=68\pm2.8(73.24\pm1.74)$ at $1\sigma$ error. Moreover, recently Capozziello et al \cite{ref73} through an effective cosmographic construction, in the framework of $f(T)$ gravity have obtained $z_{t}=0.643^{+0.034}_{-0.030}$ at $1\sigma$ error. Note that in ref \cite{ref72} authors set $\Omega_{m}=0.315$ (from Plank (2015)) to obtain above mentioned transition redshift but when they consider this parameter as a free one they obtain $z_{t}=0.247^{+0.345}_{-0.271}$ which is not accurate value. Generally, our obtained transition redshifts, except for BAO data, are consistent with what is expected in the cosmological models with present-epoch energy budget dominated by dark energy as well as standard spatially flat $\Lambda$CDM model. 


\begin{table}[ht]
\caption{Our estimated transition redshift $z_{t}$ at 1$\sigma$ for both datasets and their joint combination.}
\centering
\setlength{\tabcolsep}{2pt}
\scalebox{0.8}{
\begin{tabular} {ccccccc}
\hline
Parameter     &  OHD (CC) & BAO & SNIa & CC+BAO+Pantheon\\[0.5ex]
\hline
\hline{\smallskip}
$z_{t}$    & $0.683^{+0.087}_{-0.21}$& $1.267^{+0.086}_{-1.0}$ & $0.711^{+0.031}_{-0.35}$& $0.706^{+0.031}_{-0.34}$&\\[0.5ex]
\hline
\end{tabular}}
\label{tab:5}
\end{table}
\section{Summary}
\label{sec:6}
Cosmography is based on the Taylor expansion of the scale factor around $z=0$. This expansion is the source of two main problems arise in this kinematic approach of the study of universe. In stead of expanding scale factor we have combined the mathematical definitions of deceleration and jerk parameters (see eqs \ref{eq2b},\ref{eq2c}) which results in a general second order differential equation for squared Hubble parameter. Although the solution of this equation could lead to a general function for Hubble parameter, but it seems to be much complicated. However, it is possible to find some reasonable solutions by considering jerk parameterization. In this paper we assumed $j(z)=j_{0}$ and found a Hubble function in terms of jerk parameter. Using this function we have reconstructed other cosmographic parameters as well as deceleration-acceleration transition redshift. It is worth mentioning that since our approach, up to eq \ref{eq7}, is totally model-independent, we can constrain any derived quantity without any doubt on the validity of the estimations. Next we recovered $\Lambda$CDM model from the obtained Hubble function. It is found that when cosmological constant is responsible for the current cosmic accelerating expansion the geometry of spacetime is necessarily should be flat. For any other dynamical (time varying) dark energy scenarios the spacetime geometry should be non-flat. Probably this approach could be used for other spacetime posses some inhomogenities. We have constrained cosmpgraphic parameters as well as transition redshift and spacetime curvature over observational Hubble data \cite{ref33}, BAO \cite{ref35} and SNIa (Pantheon compilation) \cite{ref34}, and their joint combination. Our results are in agreement with almost all other results such as 9years WMAP and Planck (2015) collaboration. 
\section*{Acknowledgment}
Authors are grateful to professor Tamara Davis for critical review of the manuscript prior to submission. We are also grateful to the anonymous referee for useful comments and suggestions.

\end{document}